An Alternative Perspective on the Robust Poisson Method for Estimating Risk or Prevalence Ratios

Auhors: Denis Talbot, Miceline Mésidor, Yohann Chiu, Marc Simard, Caroline Sirois

Affiliations : Département de médecine sociale et préventive, Université Laval, Québec, Québec, Canada (Denis Talbot, Miceline Mésidor, Marc Simard); Unité santé des populations et pratiques optimales en santé, CHU de Québec – Université Laval research center, Québec, Québec, Canada (Denis Talbot, Miceline Mésidor, Caroline Sirois); Faculté de pharmacie, Université Laval, Québec, Québec, Canada (Yohann Chiu, Caroline Sirois); Institut National de Santé Publique du Québec (Marc Simard, Caroline Sirois)

Correspondance:

Denis Talbot,

Département de médecine sociale et préventive

Faculté de médecine

Université Laval

1050, avenue de la Médecine

Pavillon Ferdinand-Vandry, room 2454

Québec (Québec) G1V 0A6, Canada

Tel. 418 656-2131 ext. 405911

Email: denis.talbot@fmed.ulaval.ca

Running head: Alternative Perspective on Robust Poisson Method

Conflict of interest: None declared.

Funding: This work was supported by a grant from the Fonds de recherche du Québec – Santé [#265385 to DT]. DT and CS were supported by a career award from the Fonds de recherche du Québec – Santé.

Data access: The data are the propriety of *Ministère de la santé et des services sociaux du Québec* and the *Régie de l'assurance maladie du Québec*. Access to these data is limited to authorized personnel of the Chronic Disease and Injury Surveillance Unit.

Posted history: This manuscript was previously posted to arXiv: doi: https://doi.org/10.48550/arXiv.2112.00547

**Abstract:** The robust Poisson method is becoming increasingly popular when estimating the association of exposures with a binary outcome. Unlike the logistic regression model, the robust Poisson method yields results that can be interpreted as risk or prevalence ratios. In addition, it does not suffer from frequent non-convergence problems like the most common implementations of maximum likelihood estimators of the log–binomial model. However, using a Poisson distribution to model a binary outcome may seem counterintuitive. Methodologic papers have often presented this as a good approximation to the more natural binomial distribution. In this paper, we provide an alternative perspective to the robust Poisson method based on the semiparametric theory. This perspective highlights that the robust Poisson method does not require assuming a Poisson distribution for the outcome. In fact, the method only assumes a log–linear relationship between the risk or prevalence of the outcome and the explanatory variables. This assumption and consequences of its violation are discussed. We also provide suggestions to reduce the risk of violating the modeling assumption. Additionally, we discuss and contrast the robust Poisson method with other approaches for estimating exposure risk or prevalence ratios.

**Introduction**

A common model for estimating the association between an exposure *A* and a binary outcome *Y* while controlling for covariates (e.g., potential confounders) *L* is using a logistic regression model of the form $\text{logit}[P(Y = 1|A, L)] = \beta_0 + A\beta_1 + L\beta_2$. For a binary exposure, the exponential of the exposure's coefficient is then interpreted as the outcome's odds ratio comparing exposed to unexposed subjects conditional on *L*. However, odds ratios have been criticized as being unintuitive, difficult to interpret, and are often misinterpreted as risk ratios[1-3]. While the odds ratio provides a good approximation to the risk ratio when the outcome is rare (for example, <10%), it may not be the case otherwise[4]. In addition, the odds ratio is a non-collapsible association measure. That is, the exposure–outcome odds ratio may depend on which (or if) covariates *L* are included in the logistic regression model, even if those covariates are unrelated to the exposure and are thus not confounders, intermediate variables, or colliders[5].

The robust Poisson regression method, also called the modified Poisson, is an alternative to the logistic regression for estimating the exposure–outcome association. This approach consists in fitting a model of the form $\log[P(Y = 1|A, L)] = \log[E(Y|A, L)] = \beta_0 + A\beta_1 + L\beta_2$ proceeding as if we specified a Poisson distribution for the outcome and using a robust (also called sandwich) variance estimator of the regression parameters[6,7]. Note that because *Y* is a binary variable, $P(Y = 1|A, L) = E(Y|A, L)$, but we use the latter expression in the following to make it explicit that we treat *Y* as if it were distributed according to a Poisson distribution. In this model, the exponential of the exposure's coefficient is interpreted as the outcome's risk (or prevalence, or proportion) ratio comparing exposed and unexposed subjects conditional on *L*.

This easy-to-implement approach thus circumvents both the interpretation and the non-collapsibility challenges that arise when using a logistic regression model. In addition, simulation studies indicate that the robust Poisson method is less prone to non-convergence problems, and is more robust to model misspecifications and outliers than the maximum likelihood estimator of a log–binomial model [8-12]. It is thus unsurprising that the popularity of the robust Poisson method is on the rise. A quick Medline search of the terms "robust Poisson" or "modified Poisson" produces 203 hits for articles published in 2020, compared to 75 hits for publications in 2015.

Despite the empirical evidence in favor of the robust Poisson method, it may seem counterintuitive to fit a model that assumes a Poisson distribution to a binary outcome. Indeed, the Poisson regression is more commonly employed to model count or rate outcomes (number of events divided by the period at risk). The theoretical rationale that was advanced when the robust Poisson method was first proposed was to avoid the convergence problems of the maximum likelihood estimator of the log–binomial regression by specifying a Poisson distribution and to use a robust variance estimator to correct for the outcome distribution misspecification [6]. Others have adopted the perspective that the robust Poisson method uses a "working" Poisson distribution [13]. The goal of this paper is to provide an alternative perspective on the robust Poisson method and show that the Poisson distribution is not necessary for the approach to be valid. We believe this result makes the robust Poisson method much more intuitively appealing. In addition, we compare the strengths and limitations of the robust Poisson method and of various alternatives. We also provide an illustration comparing some of these methods in a real-world data analysis concerning the association between a material deprivation index and the use of potentially inappropriate medication among older adults.

**The robust Poisson regression from the parametric perspective**

We first briefly review the robust Poisson regression from a parametric perspective. The robust Poisson is a model of the form $\log[E(Y|A,L)] = \mu(A, L, \beta)$ where $\mu(A, L, \beta)$ is some function of $A, L, \beta$ that is linear in $\beta$, for example $\log[E(Y|A,L)] = \beta_0 + A\beta_1 + L\beta_2 + L^2\beta_3 + AL\beta_4$. To simplify the presentation, we assume the specific model $\log[E(Y|A,L)] = \beta_0 + A\beta_1 + L\beta_2$. We also temporarily assume that observations $Y_i = 1, \dots, n$ are independent and identically distributed according to a Poisson distribution, but the model can be generalized to the case where observations are correlated (repeated measures or clustered data). This model is parametric, because the distribution of the outcome is fully characterized by the model: $Y|A, L \sim \text{Poisson}[\exp(\beta_0 + A\beta_1 + L\beta_2)]$.

The parameters of the models can be estimated using the maximum likelihood approach. The likelihood and log-likelihood of the data are respectively

$$P(Y|A,L) = \prod_{i=1}^{n} \frac{\exp(\beta_0 + A_i\beta_1 + L_i\beta_2)^{Y_i} \exp[-\exp(\beta_0 + A_i\beta_1 + L_i\beta_2)]}{Y_i!}$$

$$\log[P(Y|A,L)] = \sum_{i=1}^{n}[Y_i(\beta_0 + A_i\beta_1 + L_i\beta_2) - \exp(\beta_0 + A_i\beta_1 + L_i\beta_2) - \log(Y_i!)].$$

Finding the derivative of the log-likelihood with respect to the $\beta s$ and equating to 0, the estimator $\hat{\beta}$ is the solution to the following estimating equations

$$0 = \sum_{i=1}^{n} [Y_i - \exp(\beta_0 + A_i\beta_1 + L_i\beta_2)]$$

$$0 = \sum_{i=1}^{n} A_i[Y_i - \exp(\beta_0 + A_i\beta_1 + L_i\beta_2)] \quad \text{Eq. 1}$$

$$0 = \sum_{i=1}^{n} L_i[Y_i - \exp(\beta_0 + A_i\beta_1 + L_i\beta_2)].$$

In the case of a binary outcome, the outcome distribution is misspecified: a binary variable cannot follow a Poisson distribution. Consequently, the usual asymptotic variance estimator is incorrect. Therefore, a variance estimator that is robust to this misspecification is employed: the so-called robust or sandwich variance estimator [7].

**The robust Poisson regression from the semiparametric perspective**

We now revisit the log–linear model $\log[E(Y|A,L)] = \beta_0 + A\beta_1 + L\beta_2$ without making further assumptions about the distribution of the outcome than the aforementioned log–linear relation, that is, ignoring that Y is a binary variable. This model is said to be semiparametric because a parametric assumption is made about the relationship between the mean of the outcome and the independent variables, but the remainder of the outcome distribution is left unspecified (nonparametric). Note that the results that follow are a special case of those presented in Tsiatis (2007), Section 4.1 [14]. The technical details are relegated to eAppendix 1.

Because we do not assume a specific distribution to the outcome, the usual maximum likelihood method cannot be used to estimate the parameters of the model. Instead, we can directly try to find estimating equations that produce a consistent estimator $\hat{\beta}$, that is, an estimator $\hat{\beta}$ that converges to the true value $\beta$ for large sample size. Multiple solutions to this problem are possible. Importantly, for a solution to be valid, there must first be as many independent estimating equations as there are parameters to estimate (the system of equations must be nonsingular). This condition allows each parameter to be expressed as a unique function of the observed data; a concept known as identifiability. Second, the expectation of the estimating equations must be 0. Estimating equations with expectation equal 0 are said to be unbiased and allow the consistent estimation of the parameters. We can show that a solution meeting these two conditions can be obtained by using estimating equations of the form $\sum_{i=1}^{n} M_i[Y_i - \exp(\beta_0 + A_i\beta_1 + L_i\beta_2)] = 0$, where $M_i$ is a vector with as many rows as there are parameters to be estimated (the number of $\beta$s), suitably chosen to avoid singularity of the system of equations. There is thus one estimating equation for each row of $M_i$, which results in having as many equations as there are unknowns to estimate. Moreover, assuming the log–linear model is correctly specified (i.e., assuming the outcome expectation truly is $\exp(\beta_0 + A\beta_1 + L\beta_2)$), we can show that the expectation of the estimating equations is 0, regardless of the choice of $M_i$:

$$E_{A,L,Y}\{M[Y - \exp(\beta_0 + A\beta_1 + L\beta_2)]\} = E_{A,L}(E_Y\{M[Y - \exp(\beta_0 + A\beta_1 + L\beta_2)]|A, L\})$$

$$= E_{A,L}(M\{E[Y|A, L] - \exp(\beta_0 + A\beta_1 + L\beta_2)\})$$

$$= E_{A,L}(M\{\exp(\beta_0 + A\beta_1 + L\beta_2) - \exp(\beta_0 + A\beta_1 + L\beta_2)\}) = 0,$$

where the first equality is obtained by using the law of total expectations. Consequently, any choice of $M_i$ in $\sum_{i=1}^{n} M_i[Y_i - \exp(\beta_0 + A_i\beta_1 + L_i\beta_2)] = 0$ that ensures that the equations are independent (non-singular) results in a consistent estimator. One such choice is to set the rows of $M_i$ to $(1, A_i, L_i)$. This choice yields the same estimating equations as those of the maximum likelihood when assuming a Poisson distribution (see Eq. 1), but was derived without making this assumption. Finally, a large sample (asymptotic) estimator of the variance of $\hat{\beta}$ produced using this semiparametric estimating equation procedure is the sandwich variance estimator.

The Figure provides a numerical example of the consistency of this semiparametric estimator with simulated data. We generated datasets of sample sizes from 10 to 2000, where $L$ and $A$ were generated as Bernoulli variables, with probability 0.5 and 0.55 respectively. The outcome variable ($Y$) was generated as a binary indicator using log–linear model with $\beta_0 = 0.5$, $\beta_1 = 0.3$, and $\beta_2 = 0.9$. In this example, the true risk ratio can be determined from the data-generating equations as 1.07. As shown in the Figure, as the sample size increases, the estimated risk ratio gets closer to its true value and the confidence intervals are more precise.

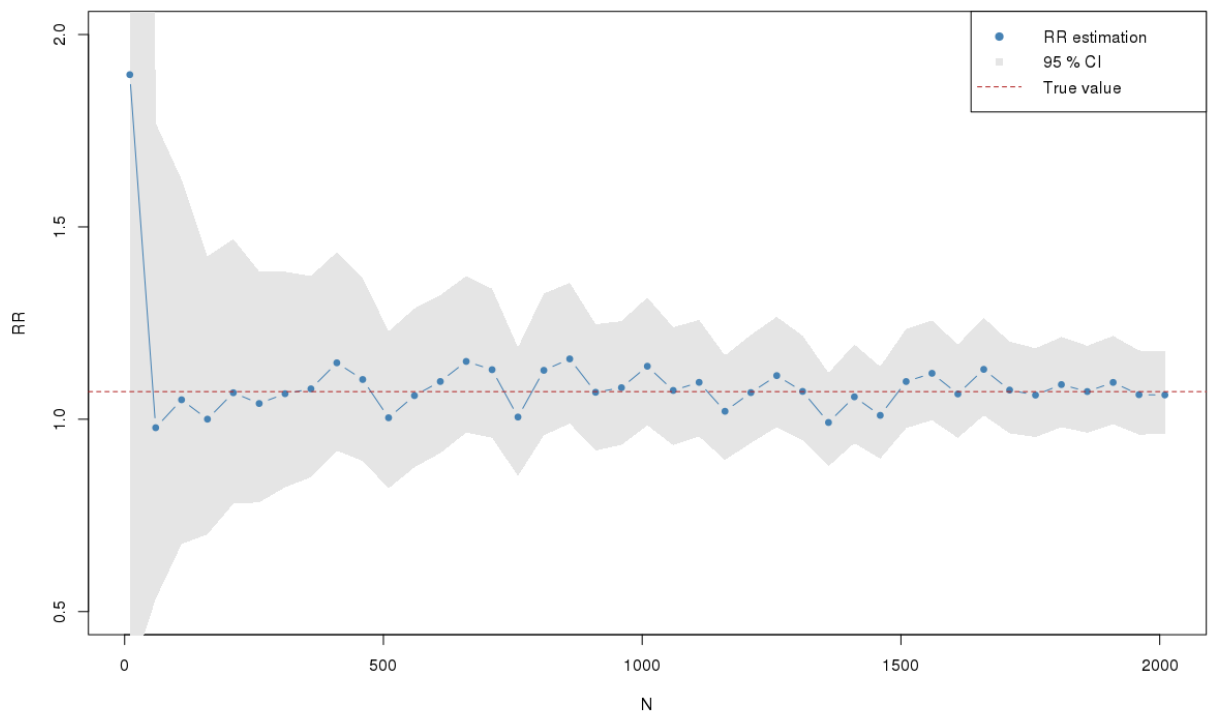

*Figure: Illustration of the consistency of the risk ratio (RR) estimator in the semiparametric log–linear model using simulated data.*

To summarize, the same estimator $\hat{\beta}$ and variance estimator can be obtained with or without making the Poisson distribution assumption for the outcome. This shows that the Poisson distribution assumption is unnecessary for the robust Poisson method to be valid. Note that this connection between parametric and semiparametric estimators is not unique to the log–linear model. For instance, for the linear regression $Y = X\beta + \varepsilon$, both the maximum likelihood estimator that assumes a normal distribution of the error (parametric) and the ordinary least squares estimator that makes no distributional assumption other than the functional form of the model (semiparametric) yield the consistent estimator $\hat{\beta} = (X^\top X)^{-1} X^\top Y$.

**A comparison of methods for estimating risk or prevalence ratios**

In this section, we provide a brief overview of alternative methods for estimating relative risks or prevalence ratios. A more in-depth review and comparison of some of these methods is available elsewhere [15].

From the semiparametric perspective that we have presented, we can see that the maximum likelihood estimator of a log–binomial model and the one obtained with a robust Poisson method are simply two different estimators of the same model. An advantage of the robust Poisson method is that it is less prone to convergence issues than traditional maximum likelihood algorithms. However, various algorithms have been introduced to avoid such convergence problems [16,17]. The R package *logbin* implements some of these algorithms [18]. A Bayesian approach has also been proposed to estimate the parameters of the log–binomial model while avoiding convergence problems [19]. One of the limitations of the robust Poisson method is that it is not an efficient estimator of the parameters of the log–binomial model [20]. This means that the robust Poisson method may result in increased variance, wider confidence intervals, and lower statistical power compared to the maximum likelihood estimator. In addition, it is possible to obtain predicted probabilities greater than 1 when using the robust Poisson method. While this property makes the robust Poisson method unsuitable for predictive purposes, various simulation results suggest that association estimates remain relatively unbiased and confidence intervals maintain their expected coverage even when implausible predicted values are produced [19,21]. Both the robust Poisson method and the maximum likelihood estimator also share the limitation of assuming that the relation between $\log[E(Y|A,L)]$ and $(A, L)$ is correctly specified in order to be consistent. The expected functional relationship between covariates and outcome, including the presence of statistical interaction terms, should be carefully considered based on subject-matter knowledge to minimize misspecifications, and thus residual confounding bias. Modeling continuous confounders in a data-adaptive way, using polynomial terms or splines for example, can also be considered to reduce misspecifications [22]. Whenever the sample size allows it, we recommend favoring rich models featuring interactions and polynomial or spline terms. However, as argued by [23], the log–linear model for a binary outcome is assuredly misspecified whenever the range of the covariates $(A,L)$ is unbounded. As such, the best one may hope to achieve by using a rich model is to limit the residual confounding bias due to misspecifications, but never avoid it completely.

More sophisticated alternatives with various advantages have also been proposed to estimate risk or prevalence ratios. For example, Fitzmaurice et al introduced a nearly efficient estimating equation method, that is, whose variance is close to the lowest possible bound [20].

Vansteelandt and Dukes developed a general approach to define main effect estimands that reduce to parameters in generalized linear models when these models are correctly specified and that still capture the exposure–outcome association when the models are misspecified [24]. This approach can further be combined with machine-learning procedures to minimize the statistical assumptions required for estimation without affecting ease of interpretation [24]. A two-step procedure was put forward by Tchetgen Tchetgen to obtain an efficient semi-parametric estimator of the relative risk [25]. This author also developed a double robust estimator for the special case where interest lies in the association of a single binary exposure variable with the outcome [25]. In this context, double robust estimators require modeling both the outcome and the exposure and are consistent if either model is correct, but not necessarily both.

In the specific setting of a single binary exposure, various other double robust estimators are also available, such as augmented inverse probability weighting and targeted maximum likelihood (TMLE) that estimate a marginal exposure risk ratio and are implemented in the R packages *AIPW* and *tmle*, respectively [26-28]. Both packages support using machine learning algorithms to minimize modeling assumptions. An R function *tmle_multi.R* is available in [29] to implement the targeted maximum likelihood algorithm with a multilevel exposure. An important consideration when using machine learning algorithms together with double robust approaches is the rate of convergence of these methods. If the machine learning methods used to model the exposure and outcome converge too slowly toward the true models as a function of sample size, the usual statistical properties of the estimators, such as asymptotic normal distribution and variance that decreases at a rate *1/n*, are no longer guaranteed. More formally, the asymptotic normality and the $\sqrt{n}$-consistency of estimators of the risk ratio estimator is only guaranteed if the product of the error terms of the nuisance models is $o(n^{-1/2})$ (for example, see reference [30] and references therein).

Another doubly robust method for situations where a single binary exposure is of interest is the binary regression model of Richardson et al [23]. This approach estimates the exposure conditional risk ratio and allows this risk ratio to vary according to the covariates' values. This approach solves multiple problems of the robust Poisson approach: it yields predicted probabilities that are constrained between 0 and 1, the model is not necessarily misspecified whenever the range of (A, L) is unbounded, and its double robustness property increases the chance of obtaining valid estimates. The fact that its estimated risk ratio varies according to the covariates' values is theoretically necessary for a model for the relative risk to be correctly specified.

Despite their strengths, many of these more sophisticated methods may be less accessible than the robust Poisson to most analysts because of their more advanced mathematical presentation or, in some cases, the lack of simple software to implement them.

We present a small-scale simulation study in eAppendix 2 comparing maximum likelihood estimators of the log–binomial model, the robust Poisson approach, the binary regression model of Richardson et al and TMLE. Each method was either implemented using a simple specification (including only main terms for the covariates) or using a rich specification (including spline terms and interactions between covariates for most methods and machine learning approaches for TMLE). Three scenarios with varying degree of complexity of the relation between covariates and both exposure and outcome were considered. For all methods, using a rich specification markedly reduced the bias in more complex scenarios. In particular, the robust Poisson approach produced estimates with negligible bias and appropriate coverage of confidence intervals in all scenarios we considered, thus performing slightly better than the more sophisticated alternatives of the binary regression model and TMLE in the more complex scenarios.

**Data illustration**

We now contrast the robust Poisson method to some alternatives in a real-data analysis. This illustration concerns the association between material deprivation and the use of potentially inappropriate medications among community-dwelling older adults in Quebec, Canada. It is a reanalysis of a subset of the data considered in reference [31].

Briefly, potentially inappropriate medications are drugs with greater risks than benefits among older adults and that should thus be avoided in this population [32]. Using data from the Quebec Integrated Chronic Disease Surveillance System database, we took a random sample of 100,000 individuals aged 66 and older on 1 April 2014. To be included in the sample, individuals had to be continuously covered by the public drug insurance plan between 1 April 2014 and 31 March 2016, be alive on 31 March 2016, and have no missing data on any of the variables of interest. Those who were transferred to a long-term care facility during the study period were excluded as their medication use is no longer captured in the database. The outcome was the use of any potentially inappropriate medication (yes vs no) between 1 April 2015 and 31 March 2016, as determined using the 2015 version of Beers criteria [33]. The exposure was an index of material deprivation (a validated ecologic proxy for socioeconomic status [34]) divided into quintiles. Covariates included age, sex, a social deprivation index [34] and the number of chronic diseases among hypertension, cardiovascular diseases, osteoporosis, diabetes, respiratory diseases, mental disorders, and Alzheimer's disease and related disorders, all measured on 1 April 2015. More information on the data is available elsewhere [31].

To estimate the association between material deprivation and the use of potentially inappropriate medications, we used 1) the robust Poisson method, 2) a maximum likelihood log–binomial method, and 3) a targeted maximum likelihood estimator (TMLE). We chose these methods because R code that supported a multilevel exposure was available. For the robust Poisson and the log–binomial methods, we modeled age using a restricted cubic spline. We considered two different maximum likelihood algorithms: the adaptive-barrier and the expectation-maximization algorithm. For TMLE, we modeled exposure using a polychotomous regression and multiple classification method [35] and the outcome using the Super Learner [36] with a logistic regression, a generalized additive model, and random forests as learners.

Descriptive statistics are provided in Table 1. The mean age was 75 years, 55% of individuals were women, subjects were close to evenly distributed among material and social deprivation quintiles, and only 14% had no chronic disease. Table 2 provides the prevalence ratio estimates comparing the different material deprivation quintiles to the lowest quintile and their 95% confidence intervals. Most methods yielded similar results, except the expectation-maximization maximum likelihood algorithm whose estimates appear to be incorrect (all prevalence ratios = 1) and failed to produce standard errors. The results of the other methods indicate that greater material deprivation is associated with a greater prevalence of using potentially inappropriate medications, which was also observed in reference [31]. The 95% confidence intervals were overall slightly wider for the robust Poisson method than for the maximum likelihood log–binomial method, which were themselves slightly wider than for TMLE.

Table 1: Baseline characteristics for a random sample of 100,000 community-dwelling older adults in Quebec, Canada

| | |
|---|---|
| Age – mean (standard deviation) | 75.0 (6.9) |
| Woman – n (%) | 55,472 (55) |
| Material deprivation quintiles – n (%) | |
| - 1 (lowest) | 18,417 (18) |
| - 2 | 19,226 (19) |
| - 3 | 20,076 (20) |
| - 4 | 21,371 (21) |
| - 5 (highest) | 20,910 (21) |
| Social deprivation quintiles – n (%) | |
| - 1 (lowest) | 17,278 (17) |
| - 2 | 18,864 (19) |
| - 3 | 21,108 (21) |
| - 4 | 21,343 (21) |
| - 5 (highest) | 21,407 (21) |
| Number of chronic diseases[a] – n (%) | |
| - 0 | 14,208 (14) |
| - 1 | 27,752 (28) |
| - 2 | 25,423 (25) |
| - 3 | 15,883 (16) |
| - ≥4 | 16,734 (17) |

[a]The chronic diseases considered were hypertension, cardiovascular diseases, osteoporosis, diabetes, respiratory diseases, mental disorders, and Alzheimer's disease and related disorders

Table 2: Association between material deprivation quintiles and use of potentially inappropriate medications in a sample of 100,000 community-dwelling older adults of Quebec, Canada

|  | Robust Poisson | Log–binomial (AB) | Log–binomial (EM)[a] | TMLE |
|---|---|---|---|---|
|  | PR (95% CI) | PR (95% CI) | PR (95% CI) | PR (95% CI) |
| Material deprivation |  |  |  |  |
| - 1 (lowest) | 1.00 | 1.00 | 1.00 | 1.00 |
| - 2 | 1.05 (1.03, 1.07) | 1.05 (1.02, 1.07) | 1.00 (NA, NA) | 1.05 (1.04, 1.06) |
| - 3 | 1.08 (1.05, 1.10) | 1.07 (1.05, 1.09) | 1.00 (NA, NA) | 1.08 (1.06, 1.09) |
| - 4 | 1.09 (1.06, 1.11) | 1.08 (1.06, 1.10) | 1.00 (NA, NA) | 1.09 (1.08, 1.10) |
| - 5 (highest) | 1.10 (1.08, 1.13) | 1.10 (1.08, 1.12) | 1.00 (NA, NA) | 1.10 (1.09, 1.12) |

All analyses were adjusted for age, sex, social deprivation and number of chronic diseases. AB = adaptive-barrier, EM = expectation-maximization, TMLE = targeted maximum likelihood estimator, PR = prevalence ratios, CI = confidence intervals.

[a]The EM estimator of the log–binomial model failed to produce standard errors

**Discussion**

Although the robust Poisson regression is an increasingly popular approach to estimate exposure effects on a risk ratio scale, the approach is frequently misunderstood. The name of the approach suggests that it is necessary to assume a Poisson distribution for the outcome, which would be inappropriate for modeling a binary outcome. In this paper, we have shown that the robust Poisson regression is a semiparametric estimator of the log–binomial model which makes no assumption on the outcome distribution except for the functional form relating the outcome probability to the covariates. As such, we have shown that it is unnecessary to make a Poisson distribution assumption for the outcome when using the robust Poisson regression. To avoid confusion on the Poisson distribution assumption, authors may prefer referring to the robust Poisson regression as a semiparametric log–linear model instead. We emphasize that the results we have presented are not new per se. However, because the mathematical theory behind them is intricate, we believe many analysts will benefit from our exposition, which we aimed to make as accessible as possible.

We have also discussed the strengths and limitations of the robust Poisson method as compared to alternatives. The main strengths of the robust Poisson method are its ease of implementation using common statistical software, its relative simplicity, its general applicability, and being less prone to convergence problems than traditional maximum likelihood estimators. Various more sophisticated alternatives have been developed and have important benefits over the robust Poisson method, including reduced variance and lower risk of being misspecified, and are thus theoretically preferable. However, for these methods to become more widely adopted by epidemiologists, we believe there is a need for better knowledge translation of these methods in a non-mathematical language, and to develop easy to use and well-documented statistical software. In our real-data illustration, the robust Poisson method, maximum likelihood estimation of the log–binomial model, and TMLE all produced similar results, both in terms of point estimates and width of confidence intervals. In conclusion, we hope this paper will be helpful to analysts to better appreciate the strengths, limitations, and assumptions of the robust Poisson method for estimating risk and prevalence ratios.

# Appendix for An Alternative Perspective on the Robust Poisson Method for Estimating Risk or Prevalence Ratios

## eAppendix 1

In this Appendix, we sketch a demonstration of the consistency of the semiparametric estimator of the regression coefficients of the log-linear model for a binary outcome under the assumption that the log-linear model is correctly specified. We also compute the asymptotic variance of the estimator and show that it can be consistently estimated by the same variance estimator as the one of the robust Poisson regression. Together, these results demonstrate the equivalence between the parametric robust Poisson regression and the semiparametric log-linear model. As a consequence, these results show that it is not necessary to assume that the outcome follows a Poisson distribution when using the robust Poisson regression. Our demonstration is based on Tsiatis (2007) Sections 3.2 and 4.1, as well as van der Vaart (2000) Chapter 5.

**Coefficient estimation**

To simplify the presentation, we denote $m(X, \beta) = M[Y - \exp(\beta_0 + A\beta_1 + L\beta_2)]$. For the estimator $\hat{\beta}$ to be consistent, $M$ must be chosen such that

$$\mathbb{E}\left\{\frac{\partial m(X, \beta)}{\partial \beta^\top}\right\} \tag{1}$$

is nonsingular, and

$$\frac{1}{n}\sum_{i=1}^{n}\left\{\frac{\partial m(X_i, \beta^*)}{\partial \beta^\top}\right\} \xrightarrow{P} \mathbb{E}\left\{\frac{\partial m(X, \beta^*)}{\partial \beta^\top}\right\} \tag{2}$$

uniformly for $\beta^*$ in a neighborhood of the true value $\beta$. First, we can show that $\mathbb{E}[m(X, \beta)] = \mathbb{E}\{[M(Y - \exp(\beta_0 + A\beta_1 + L\beta_2))]\} = 0$ if the model is correctly specified. Indeed

$$\begin{aligned}
\mathbb{E}\{M[Y - \exp(\beta_0 + A\beta_1 + L\beta_2)]\} &= \mathbb{E}_{(A,L)}\left(\mathbb{E}\left\{M[Y - \exp(\beta_0 + A\beta_1 + L\beta_2)]|A, L\right\}\right) \\
&= \mathbb{E}_{(A,L)}\left\{M\left[\mathbb{E}[Y|A, L] - \exp(\beta_0 + A\beta_1 + L\beta_2)\right]\right\} \\
&= \mathbb{E}_{(A,L)}\left\{M\left[\exp(\beta_0 + A\beta_1 + L\beta_2) - \exp(\beta_0 + A\beta_1 + L\beta_2)\right]\right\} \\
&= 0,
\end{aligned}$$

where the second equality is obtained because $M$ is a constant conditional on $A$ and $L$. As a consequence, the true parameter $\beta$ is identified from the observed data as the unique solution to $\mathbb{E}[m(X, \beta)] = 0$.

Moreover, by the uniform law of large numbers

$$\frac{1}{n}\sum_{i=1}^{n} m(X_i, \beta^*) \xrightarrow{P} \mathbb{E}\left[m(X, \beta^*)\right]$$

uniformly for all $\beta^*$ in some compact neighborhood of $\beta$. Because of the uniform convergence, and by continuity, this implies that $\hat{\beta}$ that solves the empirical estimating equations $\frac{1}{n}\sum_{i=1}^{n} m(X_i, \hat{\beta}) = 0$ converges to the true value $\beta$ that solves the population estimating equations $\mathbb{E}[m(X, \beta)] = 0$. This demonstrates the consistency of the estimator $\hat{\beta}$. A more technical presentation can be found in van der Vaart (2000) Chapter 5.

**Variance estimation**

We now turn our attention to the estimation of the variance of the estimator $\hat{\beta}$. According to the mean value theorem, if $f$ is a continuous and differentiable function over an interval $[a, b]$, then there is $c \in [a, b]$ such that

$$f'(c) = \frac{f(b) - f(a)}{b - a},$$

where $f'$ is the derivative of $f$. This mean value theorem allows writing $f(b) = f(a) + f'(c)(b - a)$ which is called a mean value expansion. We now perform a mean value expansion of $f(\hat{\beta}) = \frac{1}{n} \sum_{i=1}^{n} m(X_i, \hat{\beta})$ using $\beta^*$ as some intermediate value between $\hat{\beta}$ and $\beta$:

$$\frac{1}{n} \sum_{i=1}^{n} m(X_i, \hat{\beta}) = \frac{1}{n} \sum_{i=1}^{n} m(X_i, \beta) + \left\{ \frac{1}{n} \sum_{i=1}^{n} \frac{\partial m(X_i, \beta^*)}{\partial \beta^\top} \right\} (\hat{\beta} - \beta). \tag{3}$$

Recall that $\frac{1}{n} \sum_{i=1}^{n} m(X_i, \hat{\beta}) = 0$. As a consequence, (3) = 0. Using this in addition to the nonsingularity assumption (1), (3) can be written as

$$\sqrt{n}(\hat{\beta} - \beta) = - \left\{ \frac{1}{n} \sum_{i=1}^{n} \frac{\partial m(X_i, \beta^*)}{\partial \beta^\top} \right\}^{-1} \left[ \sqrt{n} \frac{1}{n} \sum_{i=1}^{n} m(X_i, \hat{\beta}) \right]. \tag{4}$$

Because the estimator $\hat{\beta}$ is consistent ($\hat{\beta} \xrightarrow{P} \beta$) as demonstrated earlier and because $\beta^*$ is between $\hat{\beta}$ and $\beta$, we must also have $\beta^* \xrightarrow{P} \beta$. This, in addition to the uniform convergence (2), implies that

$$\left\{ \frac{1}{n} \sum_{i=1}^{n} \frac{\partial m(X_i, \beta^*)}{\partial \beta^\top} \right\}^{-1} = \left[ \mathbb{E} \left\{ \frac{\partial m(X, \beta)}{\partial \beta^\top} \right\} \right]^{-1} + o_p(1), \tag{5}$$

where $o_p(1)$ is a term that converges to 0 as sample size increases. In other words, the estimator on the left-hand side of Equation (5) is equal to its true value plus some error that vanishes when sample size increases (right-hand side of Equation (5)). This is a different way to write that the estimator on the left-hand side of Equation (5) converges to the true value on the right-hand side of Equation (5). Inserting this result in (4) gives

$$\sqrt{n}(\hat{\beta} - \beta) = - \left[ \mathbb{E} \left\{ \frac{\partial m(X, \beta)}{\partial \beta^\top} \right\} \right]^{-1} \left[ \sqrt{n} \frac{1}{n} \sum_{i=1}^{n} m(X_i, \hat{\beta}) \right] + o_p(1). \tag{6}$$

Using the central limit theorem, we get that $\sqrt{n}(\hat{\beta} - \beta)$ converges to a normal distribution with mean

$$- \left[ \mathbb{E} \left\{ \frac{\partial m(X, \beta)}{\partial \beta^\top} \right\} \right]^{-1} \{\sqrt{n} \mathbb{E}[m(X, \beta)]\} = 0$$

since $\mathbb{E}[m(X, \beta)] = 0$ as we have shown previously. To compute the asymptotic variance of $\sqrt{n}(\hat{\beta} - \beta)$, first remark that $- \left[ \mathbb{E} \left\{ \frac{\partial m(X, \beta)}{\partial \beta^\top} \right\} \right]^{-1}$ is fixed. Noting that if $B$ is a fixed matrix, then $var(BX) =$ $Bvar(X)B^\top$, we get the following expression for the asymptotic variance of $\sqrt{n}(\hat{\beta} - \beta)$

$$\left[\mathbb{E}\left\{\frac{\partial m(X,\beta)}{\partial \beta^\top}\right\}\right]^{-1} var\{m(X,\beta)\} \left[\mathbb{E}\left\{\frac{\partial m(X,\beta)}{\partial \beta^\top}\right\}\right]^{-1^\top}.$$

Combining these results together, we get

$$\sqrt{n}(\hat{\beta} - \beta) \xrightarrow{D} N\left(0, \left[\mathbb{E}\left\{\frac{\partial m(X,\beta)}{\partial \beta^\top}\right\}\right]^{-1} var\{m(X,\beta)\} \left[\mathbb{E}\left\{\frac{\partial m(X,\beta)}{\partial \beta^\top}\right\}\right]^{-1^\top}\right).$$

An asymptotic estimator for the variance is obtained by plugging the following consistent estimators in the above formula

$$\hat{\mathbb{E}}\left\{\frac{\partial m(X,\beta)}{\partial \beta^\top}\right\} = \frac{1}{n}\sum_{i=1}^n \frac{\partial m(X_i,\hat{\beta})}{\partial \beta^\top}$$

$$\widehat{var}\{m(X,\beta)\} = \frac{1}{n}\sum_{i=1}^n m(X_i,\hat{\beta})m(X_i,\hat{\beta})^\top.$$

Because $\frac{1}{n}\sum_{i=1}^n \frac{\partial m(X_i,\hat{\beta})}{\partial \beta^\top}$ is symmetric, this yields the following variance estimator

$$\left(\frac{1}{n}\sum_{i=1}^n \frac{\partial m(X_i,\hat{\beta})}{\partial \beta^\top}\right)^{-1} \left(\frac{1}{n}\sum_{i=1}^n m(X_i,\hat{\beta})m(X_i,\hat{\beta})^\top\right) \left(\frac{1}{n}\sum_{i=1}^n \frac{\partial m(X_i,\hat{\beta})}{\partial \beta^\top}\right)^{-1}. \quad (7)$$

**Specific case of the robust Poisson method**

This result holds quite generally, for various choices of $M_i$. For the specific choice $M_i = (1, A_i, L_i^\top)^\top$, the estimating equations $\sum_{i=1}^n m(X_i, \hat{\beta}) = 0$ are

$$\sum_{i=1}^n [Y_i - \exp(\beta_0 + A_i\beta_1 + L_i\beta_2)] = 0$$

$$\sum_{i=1}^n A_i[Y_i - \exp(\beta_0 + A_i\beta_1 + L_i\beta_2)] = 0$$

$$\sum_{i=1}^n L_i[Y_i - \exp(\beta_0 + A_i\beta_1 + L_i\beta_2)] = 0,$$

which is the same as the estimating equations of the robust Poisson regression. This implies that the estimator $\hat{\beta}$ of the log-binomial model is the same as the one of the robust Poisson regression. Following Liang and Zeger (1986), the variance estimator of $\hat{\beta}$ in the robust Poisson regression is given by

$$n\left(\sum_{i=1}^n \frac{\partial \hat{\mu}_i}{\partial \beta}^\top \hat{\mu}_i^{-1} \frac{\partial \hat{\mu}_i}{\partial \beta}\right)^{-1} \left(\sum_{i=1}^n \frac{\partial \hat{\mu}_i}{\partial \beta}^\top \hat{\mu}_i^{-1}(Y_i - \hat{\mu}_i)(Y_i - \hat{\mu}_i)^\top \hat{\mu}_i^{-1} \frac{\partial \hat{\mu}_i}{\partial \beta}\right) \left(\sum_{i=1}^n \frac{\partial \hat{\mu}_i}{\partial \beta}^\top \hat{\mu}_i^{-1} \frac{\partial \hat{\mu}_i}{\partial \beta}\right)^{-1}, \quad (8)$$

where $\hat{\mu}_i = \exp(\hat{\beta}_0 + A_i\hat{\beta}_1 + L_i\hat{\beta}_2)$. We note that

$$\frac{\partial \hat{\mu}_i}{\partial \beta} = (M_i \exp(\hat{\beta}_0 + A_i\hat{\beta}_1 + L_i\hat{\beta}_2))^\top = M_i^\top \hat{\mu}_i. \tag{9}$$

Inserting (9) in (8) yields

$$n \left(\sum_{i=1}^n M_i \hat{\mu}_i \hat{\mu}_i^{-1} M_i^\top \hat{\mu}_i\right)^{-1} \left(\sum_{i=1}^n M_i \hat{\mu}_i \hat{\mu}_i^{-1} (Y_i - \hat{\mu}_i)(Y_i - \hat{\mu}_i)^\top \hat{\mu}_i^{-1} M_i^\top \hat{\mu}_i\right) \left(\sum_{i=1}^n M_i \hat{\mu}_i \hat{\mu}_i^{-1} M_i^\top \hat{\mu}_i\right)^{-1}$$

$$= n \left(\sum_{i=1}^n M_i M_i^\top \hat{\mu}_i\right)^{-1} \left(\sum_{i=1}^n M_i (Y_i - \hat{\mu}_i)(Y_i - \hat{\mu}_i)^\top M_i^\top\right) \left(\sum_{i=1}^n M_i M_i^\top \hat{\mu}_i\right)^{-1}. \tag{10}$$

Next, remark that

$$\frac{\partial m(X_i, \hat{\beta})}{\partial \beta^\top} = M_i M_i^\top \hat{\mu}_i \tag{11}$$

$$M_i(Y_i - \hat{\mu}_i) = m(X_i, \hat{\beta}). \tag{12}$$

Using (11) and (12), we can write (10) as

$$n \left(\sum_{i=1}^n \frac{\partial m(X_i, \hat{\beta})}{\partial \beta^\top}\right)^{-1} \left(\sum_{i=1}^n m(X_i, \hat{\beta}) m(X_i, \hat{\beta})^\top\right) \left(\sum_{i=1}^n \frac{\partial m(X_i, \hat{\beta})}{\partial \beta^\top}\right)^{-1}. \tag{13}$$

Because $\left(\sum_{i=1}^n \frac{\partial m(X_i, \hat{\beta})}{\partial \beta^\top}\right)$ is symmetric,

$$\left(\sum_{i=1}^n \frac{\partial m(X_i, \hat{\beta})}{\partial \beta^\top}\right)^{-1} = \left(\sum_{i=1}^n \frac{\partial m(X_i, \hat{\beta})}{\partial \beta^\top}\right)^{-1^\top}. \tag{14}$$

Using (14) in (13) and inserting the sample size $n$ within the parentheses, we find that the estimator of the variance of $\hat{\beta}$ in the robust Poisson regression can be written as

$$\left(\frac{1}{n}\sum_{i=1}^n \frac{\partial m(X_i, \hat{\beta})}{\partial \beta^\top}\right)^{-1} \left(\frac{1}{n}\sum_{i=1}^n m(X_i, \hat{\beta}) m(X_i, \hat{\beta})^\top\right) \left(\frac{1}{n}\sum_{i=1}^n \frac{\partial m(X_i, \hat{\beta})}{\partial \beta^\top}\right)^{-1},$$

which is the same expression as the estimator of the variance of $\hat{\beta}$ in the semiparametric log-linear model (see Equation (7)). This concludes the demonstration that the semiparametric log-linear model is equivalent to the robust Poisson regression and thus that the Poisson distribution assumption is unnecessary for the robust Poisson model to be valid.

# eAppendix 2

This appendix provides details concerning a small-scale simulation study that illustrates and compares different methods for estimating risk ratios. All simulations were conducted with R version 4.2.0.

**Simulation scenarios**

Three different scenarios were considered, with a varying degree of complexity in the relation between the covariates $\boldsymbol{L}$ and both exposure $A$ and outcome $Y$. In all scenarios, two covariates ($L_1$ and $L_2$) are considered and $A$ is a binary variable to allow comparing more methods. In addition, in all scenarios, we examined the distribution of $P(A|\boldsymbol{L})$ to ensure that very few values are close to 0 or 1, since such values can cause estimation problems for methods that require modeling the exposure.

**Simple scenario** In the simple scenario, $L_1$ and $L_2$ were generated as two independent binary variables. $A$ was generated as a function of $L_1$ and $L_2$, and $Y$ as a function of $A$, $L_1$ and $L_2$, without any interaction terms between the covariates. The data-generating equations were

$$L_1 \sim Bernoulli(p = 0.5)$$
$$L_2 \sim Bernoulli(p = 0.25)$$
$$A \sim Bernoulli(p = \text{expit}(-0.2 + 0.4L_1 + 0.3L_2))$$
$$Y \sim Bernoulli(p = \text{expit}(-0.4 + 0.5A - 0.5L_1 - 0.2L_2).$$

**Moderate scenario** In the moderate scenario, $L_1$ and $L_2$ are generated as continuous variables with unbounded range of values. As a consequence, the log-binomial model is necessarily misspecified to some extent. In addition, both $A$ and $Y$ are generated as a function of polynomial terms of $L_1$ and $L_2$ as well as interaction terms between $L_1$ and $L_2$. The equations were

$$L_1 \sim N(0,1)$$
$$L_2 \sim N(0,1)$$
$$A \sim Bernoulli(p = \text{expit}(-0.2 + 0.3L_1 + 0.2L_1^2 + 0.1L_1^3 + 0.3L_2 - 0.2L_2^2$$
$$- 0.3L_1L_2 + 0.2L_1^2L_2 - 0.2L_1L_2^2))$$
$$Y \sim Bernoulli(p = \text{expit}(-0.4 + 0.5A - 0.5L_1 - 0.2L_1^2 - 0.2L_2 + 0.1L_2^2 + 0.1L_2^3 + 0.5L_1L_2)).$$

**Complex scenario** As in the moderate scenario, $L_1$ and $L_2$ are generated as continuous variables with unbounded range of values in the complex scenario. In addition, $A$ and $Y$ are generated as a function of non-polynomial transformations of $L_1$ and $L_2$. The equations were

$$L_1 \sim N(0,1)$$
$$L_2 \sim N(0,1)$$
$$A \sim Bernoulli(p = \text{expit}(-0.2 + 2\sin L_1 + |\sin L_2| - 0.3|L_1|\cos L_2))$$
$$Y \sim Bernoulli(p = \text{expit}(-0.4 + 0.5A - 2\sin L_1 - |L_2| + |L_1|\sin L_2).$$

**Simulation analysis**

For all scenarios, we first computed the true exposure marginal risk ratio by Monte Carlo simulation by generating $n = 1,000,000$ observations of $\boldsymbol{L}$ and then generating counterfactual outcomes under exposure ($Y^1$) and no exposure ($Y^0$) for each of these observations. The true risk ratios were 1.35 in the simple scenario, 1.28 in the moderate scenario and 1.25 in the complex scenario. We then generated 1000 independent samples of size $n = 1000$ according to the equations presented above.

The risk ratio was estimated using the usual maximum likelihood log-binomial approach (ML) implemented in the `glm` function, a log-binomial model with an expectation-maximization algorithm (EM), a log-binomial model with an adaptive-barrier algorithm (AB), the robust Poisson approach, the double robust binary regression model of Richardson et al (2017), and targeted maximum likelihood estimation (TMLE). For each approach, we considered both a simple specification and a rich specification. For the simple specification, only main terms of the covariates were included. In the rich specification, continuous covariates were modeled using restricted cubic splines (using the default options of the `rcs` function from the `rms` package) and including a $L_1 \times L_2$ term for most methods, except for TMLE where the rich model consisted in using a Super Learner with generalized linear models, generalized additive models and random forests as learners.

For each method, we computed the empirical bias (difference between the average of the estimates and the true value), root mean squared error (RMSE, the sum of the square of the bias and the empirical variance) and the coverage of 95% confidence intervals (the proportion of the replications where the 95% confidence interval included the true value).

**Simulation results**

The simulation results are presented in Table 1. The maximum Monte Carlo standard error was 0.005 for estimating bias, and 1.6 for the coverage of confidence intervals (Moris et al, 2017).

In the simple scenario, all methods produced estimates with essentially no bias, identical RMSE and 95% coverage of confidence intervals, both using the simple and rich specification.

In the moderate scenario, the log-binomial with either the usual maximum likelihood algorithm and with the adaptive-barrier algorithm failed to produce results. The log-binomial model with the EM algorithm failed to produce standard errors in 149 replications when using a simple specification and in all replications when using the rich specification. All methods produced moderately biased estimates ($\approx -0.10$) when considering a simple specification and coverage was moderately inferior to the expected 95% ($\approx 80\%$). Using a rich specification did not improve the bias of the log-binomial method with the EM algorithm. However, the bias was considerably reduced for the binary regression method and for TMLE, and virtually eliminated for the robust Poisson approach. The coverage of confidence intervals was also very close to its expected level. The RMSE of the robust Poisson, the binary regression and TMLE were similar.

In the complex scenario, the log-binomial with the usual maximum likelihood algorithm and with the adaptive-barrier algorithms again failed to produce results, and the EM algorithm failed to produce standard errors. When using a simple specification, all methods had substantial bias ($\approx -0.30$) and low coverage of confidence intervals (between 25% and 52%). Using the rich specification reduced only slightly the bias for the EM log-binomial approach (-0.27). The bias of the other approaches was much closer to 0 when using the rich specification (0.02 for the robust Poisson and -0.07 for the binary regression and TMLE) and the coverage was closer to 95% (95% for the robust Poisson, 85% for the binary regression and 82% for TMLE). The RMSE of TMLE was slightly greater than that of the robust Poisson and the binary regression model.

eTable 1: Bias, root mean squared error (RMSE) and coverage of 95% confidence intervals (CP) of different methods for estimating a risk ratio in three simulation scenarios of varying complexity

|  | | Simple | | | Moderate | | | Complex | | |
|---|---|---|---|---|---|---|---|---|---|---|
|  | Method | Bias | RMSE | CP | Bias | RMSE | CP | Bias | RMSE | CP |
| Simple specification | Log-binomial (ML) | -0.00 | 0.11 | 95 | NA | NA | NA | NA | NA | NA |
|  | Log-binomial (EM) | -0.00 | 0.11 | 95 | -0.09 | 0.12 | 81 | -0.37 | 0.38 | NA |
|  | Log-binomial (AB) | -0.00 | 0.11 | 95 | NA | NA | NA | NA | NA | NA |
|  | Robust Poisson | 0.00 | 0.11 | 95 | -0.10 | 0.13 | 81 | -0.27 | 0.29 | 25 |
|  | Binary regression | -0.00 | 0.11 | 95 | -0.08 | 0.12 | 84 | -0.24 | 0.26 | 34 |
|  | TMLE | 0.00 | 0.11 | 95 | -0.09 | 0.13 | 81 | -0.32 | 0.35 | 52 |
| Rich specification | log-binomial (ML) | -0.00 | 0.11 | 94 | NA | NA | NA | NA | NA | NA |
|  | log-binomial (EM) | -0.00 | 0.11 | 94 | -0.10 | 0.13 | NA | -0.27 | 0.28 | NA |
|  | log-binomial (AB) | -0.00 | 0.11 | 94 | NA | NA | NA | NA | NA | NA |
|  | Robust Poisson | 0.00 | 0.11 | 95 | -0.00 | 0.09 | 95 | 0.02 | 0.11 | 95 |
|  | Binary regression | -0.00 | 0.11 | 94 | -0.03 | 0.09 | 93 | -0.07 | 0.11 | 85 |
|  | TMLE | 0.00 | 0.11 | 96 | -0.04 | 0.10 | 89 | -0.07 | 0.13 | 82 |

Note: NA is reported when a method failed to yield results in most replications

**Additional investigations**

The poorer performance of the binary regression model and TMLE compared to the robust Poisson approach in the moderate and complex scenarios was an unexpected result considering the theoretical advantage of these methods over the robust Poisson. Further analyses were conducted to better understand these results. The distribution of the estimates for both approaches was symmetrical around their average and did not contain extreme values, suggesting the problem was not due the methods behaving very poorly in a few replicates. Thus, we first considered including additional learners in the Super Learner used for TMLE: generalized linear models with interactions and neural networks with a single hidden layer. The bias was only marginally reduced (-0.05) and coverage was marginally increased (86%). Further including support vector machines and earth (an implementation of multivariate adaptive regression splines) almost completely eliminated bias (-0.01) and brought coverage to an acceptable level (93%). We also tried increasing the sample size to $n = 5000$ and $n = 10,000$ and neither improved the results for the binary regression or TMLE. Finally, we tried using the maximum likelihood estimator of the binary regression model (instead of the double robust estimator) but the bias increased to $-0.14$ and coverage decreased to $0.67$.